\documentclass[twocolumn,superscriptaddress,amsmath,amssymb,aps,prl,floatfix]{revtex4-2}
\usepackage[usenames,dvipsnames]{color}
\usepackage{comment}
\usepackage{cancel}
\usepackage{ulem}
\usepackage{physics}
\usepackage{graphicx}% Include figure files
\usepackage{dcolumn}% Align table columns on decimal point
\usepackage{bm}% bold math
\usepackage{gensymb}
\usepackage{tensor}
\usepackage[percent]{overpic}
\usepackage{mathbbol}
\usepackage{multibib} %trying to compile two bibliography without errores
\usepackage[breaklinks=true,colorlinks,citecolor=blue,linkcolor=blue,urlcolor=blue]{hyperref}% add hypertext capabilities

\begin{document}
\title{Giant chirality-induced spin polarization in twisted  transition metal dichalcogenides}%
\author{Guido Menichetti}
\email{guido.menichetti@df.unipi.it}
%\altaffiliation{These authors contributed equally to this work.}
\affiliation{Dipartimento di Fisica dell'Universit\`a di Pisa, Largo Bruno Pontecorvo 3, I-56127 Pisa,~Italy}
\author{Lorenzo Cavicchi}
\affiliation{Scuola Normale Superiore, Piazza dei Cavalieri 7, I-56126 Pisa,~Italy}
\author{Leonardo Lucchesi}
%\altaffiliation{These authors contributed equally to this work.}
\affiliation{Dipartimento di Fisica dell'Universit\`a di Pisa, Largo Bruno Pontecorvo 3, I-56127 Pisa,~Italy}
\affiliation{Dipartimento di Ingegneria dell'Informazione dell'Universit\`a di Pisa, Via Girolamo Caruso 16, I-56122 Pisa,~Italy}
\author{Fabio Taddei}
\affiliation{NEST, Istituto Nanoscienze-CNR, Piazza S. Silvestro 12, I-56126 Pisa,~Italy}
\affiliation{Scuola Normale Superiore, Piazza dei Cavalieri 7, I-56126 Pisa,~Italy}
\author{Giuseppe Iannaccone}
\affiliation{Dipartimento di Ingegneria dell'Informazione dell'Universit\`a di Pisa, Via Girolamo Caruso 16, I-56122 Pisa,~Italy}
\author{Pablo Jarillo-Herrero}
\affiliation{Department of Physics, Massachusetts Institute of Technology, Cambridge, Massachusetts,~USA}
\author{Claudia Felser}
\affiliation{Max Planck Institute for Chemical Physics of Solids, N\"{o}thnitzer Str. 40, Dresden 01187,~Germany}
\author{Frank H. L. Koppens}
\affiliation{ICFO-Institut de Ci\`{e}ncies Fot\`{o}niques, The Barcelona Institute of Science and Technology, Av. Carl Friedrich Gauss 3, 08860 Castelldefels (Barcelona),~Spain}
\affiliation{ICREA-Instituci\'{o} Catalana de Recerca i Estudis Avan\c{c}ats, Passeig de Llu\'{i}s Companys 23, 08010 Barcelona,~Spain}
\author{Marco Polini}
\affiliation{Dipartimento di Fisica dell'Universit\`a di Pisa, Largo Bruno Pontecorvo 3, I-56127 Pisa,~Italy}
\affiliation{ICFO-Institut de Ci\`{e}ncies Fot\`{o}niques, The Barcelona Institute of Science and Technology, Av. Carl Friedrich Gauss 3, 08860 Castelldefels (Barcelona),~Spain}

\date{\today}% It is always \today, today,
             %  but any date may be explicitly specified

\begin{abstract}
Chirality-induced spin selectivity (CISS) is an effect that has recently attracted a great deal of attention in chiral chemistry and that remains to be understood. In the CISS effect, electrons passing through chiral molecules acquire a large degree of spin polarization. In this work we study the case of atomically-thin chiral crystals created by van der Waals assembly. We show that this effect can be spectacularly large in systems containing just two monolayers, provided they are spin-orbit coupled. Its origin stems from the combined effects of structural chirality and {\it spin-flipping} spin-orbit coupling. We present detailed calculations for twisted homobilayer transition metal dichalcogenides, showing that the chirality-induced spin polarization can be giant, e.g.~easily exceeding $50\%$ for ${\rm MoTe}_2$. Our results clearly indicate that twisted quantum materials can operate as a fully tunable platform for the study and control of the CISS effect in condensed matter physics and chiral chemistry. 
\end{abstract}
\maketitle

\begin{figure}[t]
\centering
\begin{overpic}[width=\columnwidth]{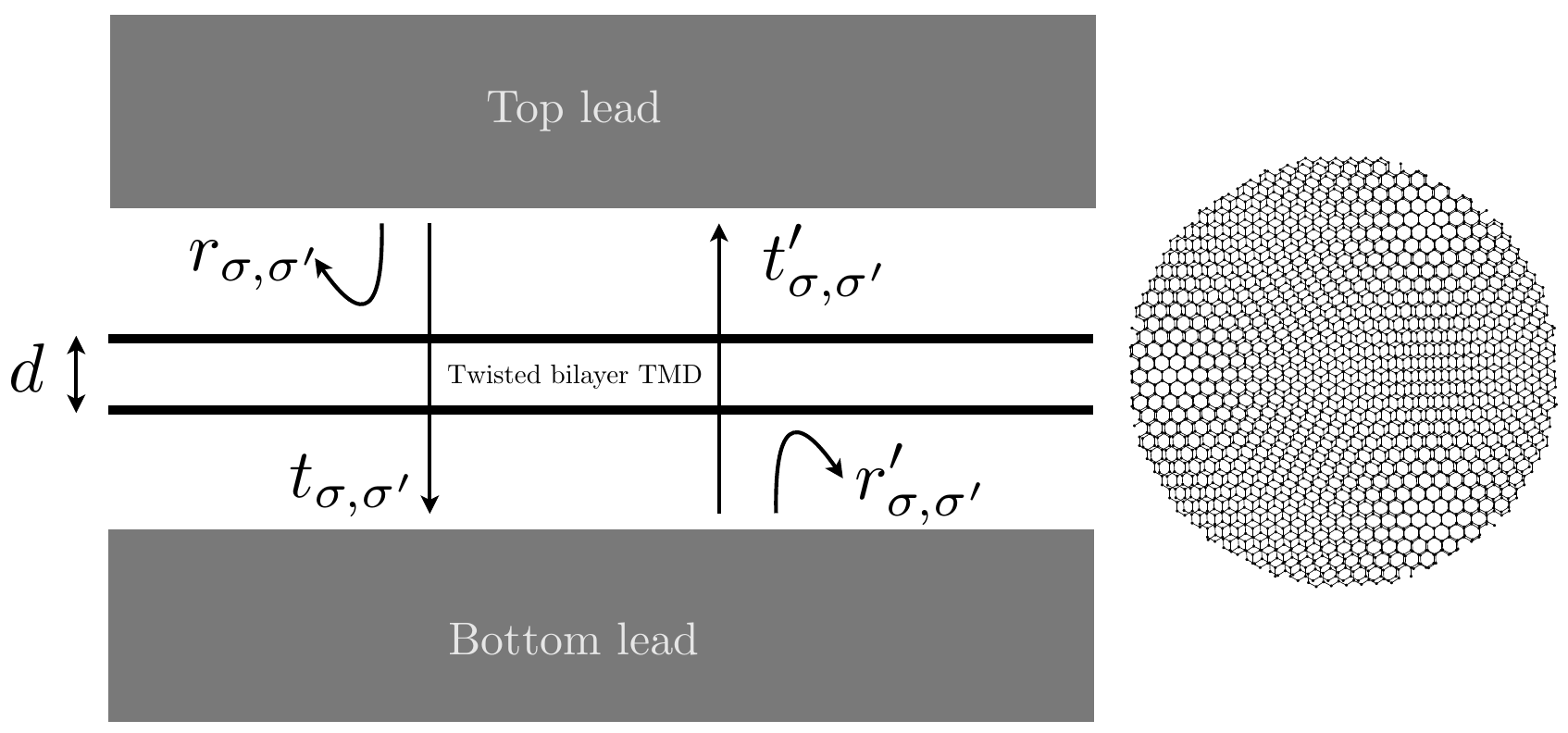} 
\end{overpic}
\caption{Sketch of the two-terminal setup we have studied in this work. A twisted homobilayer TMD, developing along the $\hat{\bm x}$-$\hat{\bm y}$ plane, is contacted by two semi-infinite leads (denoted by grey-shaded areas). The spatial separation between the two TMD layers in the vertical $\hat{\bm z}$ direction is $d$. The twist between the two layers is achieved by a counter-clockwise rotation of the top layer with respect to the bottom one by an angle $\theta$. The resulting moir\'e lattice is plotted on the right. Black arrows show the spin-resolved reflection, $r_{\sigma, \sigma^\prime}$, $r^{\prime}_{\sigma, \sigma^\prime}$, and transmission, $t_{\sigma, \sigma^\prime}$, $t^{\prime}_{\sigma, \sigma^\prime}$, components of the scattering matrix $S$ reported in Eq.~\eqref{eqn:Smatrix_def}.\label{fig1}}
\end{figure}
\begin{figure}[t]
\centering
\begin{overpic}[width=\columnwidth]{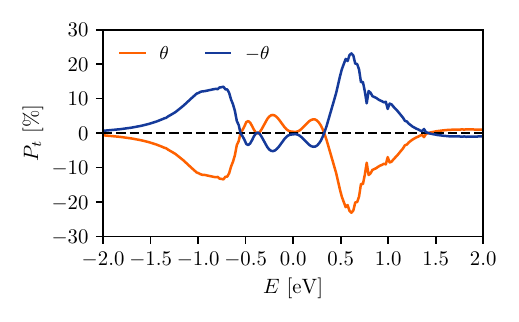} 
\end{overpic}
\caption{(Color online) Numerical results for the energy dependence of the spin polarization $P_t = P_t(E)$, as defined in Eq.~\eqref{eqn:spin_polarization}, of electrons transmitted from the top to the bottom lead. The calculated spin polarization has opposite sign for $\theta$ (orange curve) and $-\theta$ (blue curve). In the non-chiral (untwisted, $\theta=0$) case, the spin polarization $P_t$ vanishes (black dashed line). Results in this plot refer to parameters for twisted homobilayer ${\rm MoTe}_2$~\cite{SM},  $\theta = 6.01\degree$, $\lambda_0=220~{\rm meV}$, and $\lambda_{\rm BR}=0$.\label{fig2}}
\end{figure}
{\color{blue}{\it Introduction.}}---In 1999, a team led by R. Naaman recognized that films made from chiral organic molecules scatter polarized electrons asymmetrically~\cite{Ray1999}. Spin polarization induced by chirality is an effect now universally known as chirality-induced spin selectivity (CISS) effect~\cite{Naaman2012,Naaman2015,Naaman2019,Naaman2020,Yang2021}. Almost ten years later, Rosenberg et al.~\cite{Rosenberg2008} demonstrated that a polarized spin can induce an enantiospecific chemical process. It is now widely accepted~\cite{Naaman2012,Naaman2015,Naaman2019,Naaman2020,Yang2021} that the CISS effect has profound implications on asymmetric chemistry. Chiral molecules and, more in general, chiral materials~\cite{Waldeck_APL_2021} and bulk chiral crystals~\cite{Shiota2021} can act as spin polarizers and filters and therefore influence processes associated with electron transfer, electron transport, and bond polarization through chiral structures. 

To the best of our knowledge, though, a microscopic understanding of the molecular CISS effect is still lacking (for a recent summary we refer the reader to Ref.~\cite{Evers2022}). In particular, the role of spin-orbit coupling~\cite{Evers2022}, geometric phases~\cite{Liu_NatureMater_2021}, electron correlations~\cite{Fransson_JPCL_2019}, electron-phonon coupling~\cite{Fransson2020}, dissipation~\cite{Volosniev_PRB_2021}, and the interactions between chiral molecules and/or chiral and magnetic substrates are not understood. This confusion on its microscopic origin limits the use of CISS in applications and makes it challenging to enhance it.  

The point we want to raise in this work is that a thorough microscopic understanding of the CISS effect calls for a fully tunable platform where the dependence of the effect on a multitude of control parameters can be studied, both experimentally and theoretically. As we will demonstrate below, we believe that moir\'{e} superlattices obtained by twisting atomically-thin crystals with respect to each other~\cite{Andrei2020,Andrei2021,Kennes2021} represent such a platform. Despite the great deal of interest that these systems have attracted since the discovery of superconductivity~\cite{Cao2018a} and correlated insulating states~\cite{Cao2018b} in twisted bilayer graphene (TBG)~\cite{Santos2007,shallcross_prl_2008,mele_prb_2010,Li2010,shallcross_prb_2010,morell_PRB_2010,bistritzer_prb_2010,Bistritzer2011,lopes_prb_2012}, very few studies have highlighted the fact that these systems are naturally chiral materials akin to ``giant chiral molecules''. The only qualitative difference between chiral molecules and twisted quasi-two-dimensional (2D) materials is that the latter have (a nearly exact) Bloch translational invariance. 

The fact that twisted materials are chiral is not a mere mathematical statement but has experimental implications e.g.~on the interaction between these materials and light. For example, natural optical activity~\cite{Landau08}, a hallmark of chiral materials, was experimentally discovered~\cite{Kim2016} in TBG at large twist angles well before exotic states of matter were found at the magic angle~\cite{Cao2018a,Cao2018b}. TBG exhibits remarkably large circular dichroism~\cite{Kim2016,Morell_2DMaterials_2017}, up to a factor $100$ stronger than for a layer of chiral molecules of similar thickness.

In this work, we demonstrate that the interplay between spin-orbit coupling (SOC) and chirality in twisted quantum materials gives rise to a giant CISS effect, {\it even for just two twisted monolayers}. More precisely, we consider a family of twisted moir\'{e} superlattices which are known to display strong SOC. These are twisted homobilayer transition metal dichalcogenides (TMDs)~\cite{Wu_PRL_2019} such as twisted ${\rm MoTe}_2$, which is currently attracting a great deal of attention because of the experimental discovery~\cite{Cai2023,Park2023,Zeng2023,Xu2023} of fractional Chern insulating states in zero magnetic field.  These systems offer a wide range of tunable parameters, such as carrier density (which can be changed via e.g. the electrical field effect), twist angle (which can be changed at will, thereby inducing dramatic changes in the Bloch bands of these crystals), and more conventional ones, such as temperature and applied magnetic fields. Our interest here is on the study of the interplay between chiral orbital motion, spin-orbit coupling, and the spin degree of freedom. Quantifying the role of electron-electron interactions (e.g.~exchange interactions~\cite{Naaman2022}), which are strong in moir\'{e} materials, and quantum geometry~\cite{Provost1980,Xiao2010,Torma2022} is well beyond the scope of the present work and is left for future work.

{\color{blue}{\it Setup and scattering matrix approach to the CISS effect.}}---In order to numerically extract information about the CISS effect in twisted homobilayer TMDs, we utilize the setup depicted in Fig.~\ref{fig1}. It consists of a twisted homobilayer TMD contacted by two leads. One can inject carriers from the top lead, for example, which, after tunneling through the strongly spin-orbit coupled twisted homobilayer TMD, will be extracted from the bottom layer. As we will discuss momentarily, the reciprocity theorem~\cite{Buttiker1988} forbids the observation of a spin-polarized current through a standard linear-response magnetoresistance measurement in a two-terminal setup~\cite{Yang2019}. Nevertheless, the spin-polarized nature of electron scattering in our chiral van der Waals heterostructure can be detected through a spin-resolved scattering matrix approach~\cite{Nazarov,Wolf2022,Xiao2022,footnote_van_Wees}.

The top and bottom leads in Fig.~\ref{fig1} offer asymptotic propagating states. We introduce: i) $\psi^{\rm out}_{\rm T,\sigma}$ ($\psi^{\rm out}_{\rm B,\sigma}$) as the asymptotic wave-function for electrons with spin $\sigma = \uparrow, \downarrow$, which are scattered out of the twisted homobilayer TMD into the top (bottom) lead; ii) $ \psi^{\rm in}_{\rm T,\sigma^\prime}$ ($\psi^{\rm in}_{\rm B,\sigma^\prime} $) as the asymptotic wave-function for electrons with spin $\sigma^\prime = \uparrow, \downarrow$, which are injected from the top (bottom) lead into the twisted homobilayer TMD. The latter can be viewed as a scatterer, whose transport properties are completely defined by its spin-resolved scattering matrix $S_{\sigma,\sigma^\prime}$~\cite{Nazarov}:
\begin{equation}\label{eqn:Smatrix_def}
    \left(\begin{array}{c}
        \psi^{\rm out}_{\rm T,\sigma} \\
        \psi^{\rm out}_{\rm B,\sigma} 
    \end{array}\right) = \sum_{\sigma^\prime = \uparrow, \downarrow} S_{\sigma,\sigma^\prime} \left(\begin{array}{c}
        \psi^{\rm in}_{\rm T,\sigma^\prime} \\
        \psi^{\rm in}_{\rm B,\sigma^\prime} 
    \end{array}\right)~,
\end{equation}
where
\begin{equation}\label{eqn:Smatrix_def_2}
S_{\sigma, \sigma^\prime}\equiv
    \begin{pmatrix}
    r_{\sigma,\sigma^\prime} & t_{\sigma,\sigma^\prime}^\prime\\t_{\sigma,\sigma^\prime}&r_{\sigma,\sigma^\prime}^\prime
    \end{pmatrix}~.
\end{equation}
Here, $t_{\sigma,\sigma^\prime},r_{\sigma,\sigma^\prime}$ ($t_{\sigma,\sigma^\prime}^{\prime},r_{\sigma,\sigma^\prime}^{\prime}$) are the spin-resolved transmission and reflection amplitudes of electrons coming from the top (bottom) lead---see Fig.~\ref{fig1}. All the quantities in Eqs.~(\ref{eqn:Smatrix_def})-(\ref{eqn:Smatrix_def_2})---and below in Eqs.~(\ref{eq:rho})-(\ref{eqn:spin_polarization})---are functions of a single energy $E$, since we are assuming only elastic scattering mechanisms.

Starting from the entries of the scattering matrix, we introduce the following reflection $\rho_{\mathbb r},\rho_{{\mathbb r}^{\prime}}$ and transmission $\rho_{\mathbb t}, \rho_{{\mathbb t}^{\prime}}$ probability matrices:
\begin{equation}\label{eq:rho}
\begin{array}{cc}
	\rho_{\mathbb r }\equiv {\mathbb r} {\mathbb r}^\dagger~, & \rho_{{\mathbb r}^{\prime}}\equiv {\mathbb r}^{\prime} ({\mathbb r}^{\prime})^\dagger\\
    	\rho_{\mathbb t}\equiv {\mathbb t} {\mathbb t}^\dagger~, & \rho_{{\mathbb t}^{\prime}}\equiv {\mathbb t}^{\prime} ({\mathbb t}^{\prime})^\dagger~.
\end{array}
\end{equation}
Here, ${\mathbb r}$, ${\mathbb r}^{\prime}$, ${\mathbb t}$, and ${\mathbb t}^{\prime}$ are matrices in spin space with matrix elements $r_{\sigma,\sigma^\prime}$, $r_{\sigma,\sigma^\prime}^{\prime}$, $t_{\sigma,\sigma^\prime}$, $t_{\sigma,\sigma^\prime}^{\prime}$, respectively.

We also define the following dimensionless spin conductances~\cite{Wolf2022} for each scattering process:
\begin{equation}\label{eqn:spin_conductance_def}
\begin{array}{cc}
	\sigma_{r }\equiv {\rm Tr}[\hat{s}_z\rho_{\mathbb r }] ~, & \sigma_{r^\prime}\equiv {\rm Tr}[\hat{s}_z\rho_{{\mathbb r}^{\prime}}]\\
    	\sigma_{t}\equiv {\rm Tr}[\hat{s}_z\rho_{\mathbb t}]~, & \sigma_{t^\prime}\equiv {\rm Tr}[\hat{s}_z\rho_{{\mathbb t}^{\prime}}]~.
\end{array}
\end{equation}
where $\hat{s}_z$ is a Pauli matrix. The trace operation in Eq.~\eqref{eqn:spin_conductance_def} is of course intended over the spin degrees of freedom.

Once again, as demonstrated in great detail in Ref.~\cite{Yang2019}, the reciprocity theorem~\cite{Buttiker1988} allows one to conclude that any linear-response electrical measurement in a two-terminal setup, even in the presence of auxiliary ferromagnets, is {\it insensitive} to the ``spin conductances'' introduced in Eq.~(\ref{eqn:spin_conductance_def}). The setup depicted in Fig.~\ref{fig1} therefore should not be intended as a real experimental setup to infer information on the quantities in Eq.~(\ref{eqn:spin_conductance_def}), but, rather, as a computational setup with asymptotic propagating states to numerically access the full scattering matrix.

\begin{figure}[t]
\centering
\begin{tabular}{c}
\begin{overpic}[unit=1mm,width=\columnwidth]{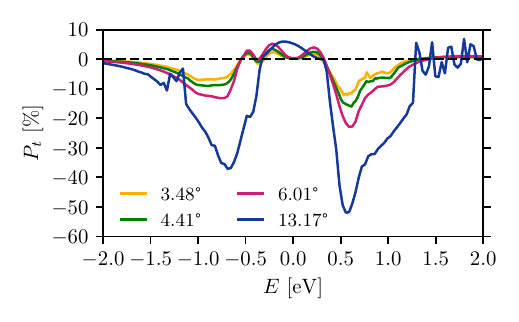} 
\put(3,56){\rm{(a)}}
\end{overpic} \\
\begin{overpic}[unit=1mm,width=\columnwidth]{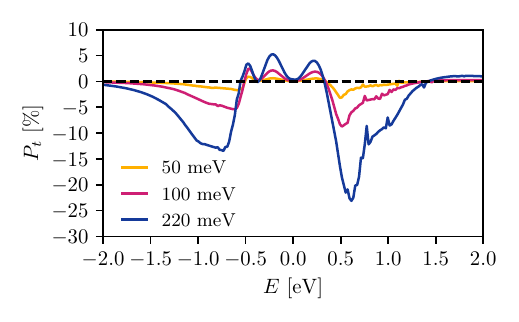} 
\put(3,56){\rm{(b)}}
\end{overpic}
\end{tabular}
\caption{(Color online) Panel (a) Results for the spin polarization $P_t(E)$ for different values of the twist angle $\theta$ ($\theta = 3.48\degree, 4.41\degree, 6.01\degree$, and $13.17\degree$). The black dashed line refers to the non-chiral $\theta = 0$ case. Panel (b) Results for the spin polarization $P_t(E)$ for different values of $\lambda_0$ ($\lambda_0 = 50~{\rm meV}, 100~{\rm meV}$, and $220~{\rm meV}$). In this panel, the angle has been fixed at $\theta=6.01^\circ$.  The black dashed line refers to the case in which SOC is artificially turned off by setting $\lambda_0 = 0$. All the other parameters are the same as in Fig.~\ref{fig2}. \label{fig3}}
\end{figure}

The four quantities in Eq.~(\ref{eqn:spin_conductance_def}) are not independent. Because of charge conservation (i.e.~because of the unitarity of the full scattering matrix), $\sigma_r=-\sigma_{t^\prime}$ and $\sigma_{r^\prime}=-\sigma_{t}$. Therefore, in general, we have two independent spin conductances, e.g.~$\sigma_t$ and $\sigma_{t^\prime}$. From now on, we focus only on $\sigma_{t}$ for reasons that will become clear momentarily. In particular,  we introduce the energy-dependent {\it chirality-induced spin polarization} $P_t(E)$ by properly normalizing the corresponding spin conductance:
\begin{equation}\label{eqn:spin_polarization}
P_t(E) \equiv \frac{\sigma_t(E)}{\Tr[\rho_{\mathbb t}(E)]} \in [-1, 1]~.
\end{equation} 
This quantity gives information on the strength of the CISS effect~\cite{Yang2019,Wolf2022} and vanishes when the scatterer is non-chiral.

{\color{blue}{\it Microscopic modelling of the scatterer.}}---For what concerns the microscopic modelling of the scatterer, i.e.~the twisted homobilayer TMD, we employed a gapped Dirac-fermion model~\cite{Wu_PRL_2019}, by introducing a mass term in a tight-binding Hamiltonian for TBG~\cite{Santos2012,Laissardiere2012}. The details of the complete tight-binding Hamiltonian are reported in Sect.~I of the Supplemental Material~\cite{SM}. Here, given its crucial importance, we focus only on the tight-binding description of SOC in twisted homobilayer TMDs. Among all the possible SOC terms allowed by $C_{3}$ symmetry~\cite{Kochan2017}, we restrict our analysis to {\it spin-flipping} nearest-neighbor intra-layer hoppings, since nearest-neighbour and next-nearest-neighbour {\it spin-conserving} hopping terms yield a vanishing contribution to the CISS strength (\ref{eqn:spin_polarization}).  The fact that the physics of CISS requires, at a general, fundamental level, spin-flip processes is in agreement with the conclusions of Ref.~\cite{Yang2019}.

The SOC Hamiltonian we consider here is then given by~\cite{Kochan2017}:
\begin{widetext}
\begin{equation}\label{eq:SOC}
	\hat{\cal H}_{\rm SOC}^{(\ell)}=\frac{2i}{3}\lambda_0^{\prime(\ell)}\sum_{\sigma\neq\sigma^\prime}\sum_{\expval{m,n}} \sum_{\tau,\tau^\prime} \left[\hat{\bm{s}}\times \bm{d}_{m,\tau,n,\tau^\prime}\right]_{\sigma,\sigma^\prime}|m, \ell, \tau\rangle\langle n, \ell, \tau^\prime|\otimes\ket{\sigma}\bra{\sigma^\prime} ~,
\end{equation}
\end{widetext}
where $\ell = 1,2$ is the layer index, $\lambda_0^{\prime(\ell)} = [\lambda_0-(-1)^{\ell}\lambda_{\rm BR}]$ is the SOC parameter (with units of energy), $\hat{\bm{s}}=(\hat{s}_x,\hat{s}_y,\hat{s}_z)$ is a vector of Pauli matrices, $\ket{m,\ell,\tau}\otimes\ket{\sigma}$ denotes the state of an electron with spin $\sigma$ localized on site $m$ and sublattice $\tau=A,B$ of layer $\ell$ (for further details, see Ref.~\cite{SM}). Finally, $\bm{d}_{m,\tau,n,\tau^\prime}$ is a dimensionless unit vector lying in the $\hat{\bm x}$-$\hat{\bm y}$ plane, which points from lattice site $n$ to the nearest-neighbor site $m$. The symbol $\langle \dots\rangle$ in one of the sums in Eq.~(\ref{eq:SOC}) refers to the fact that the sum is restricted to nearest-neighbor pairs. One important observation is now in order. The SOC parameter $\lambda_0$ is an intrinsic intra-layer term---intrinsic in the sense that is also present in Bernal-stacked bilayer graphene~\cite{Konschuh2012}. The layer-dependent term, controlled by $\lambda_{\rm BR}$, stems instead from broken inversion symmetry. This latter term, which, in an ordinary Bernal-stacked bilayer graphene is due to a perpendicular applied electric field~\cite{Konschuh2012}, here models the fact that twisted homobilayer TMDs do not have an inversion center, even in the absence of extrinsic external electric fields. Finally, we note that, in the case $\lambda_{\rm BR}=0$, our setup is invariant under the simultaneous exchange  of a) top and bottom leads and b) top and bottom layers of the twisted homobilayer TMD. In this case, one has $\sigma_{t^\prime} =- \sigma_t$. For $\lambda_{\rm BR}\neq 0$, $\sigma_{t^\prime} \neq - \sigma_t$.

While certainly approximate, this lightweight model of the scatterer treats on equal footing the structural chirality of the system (i.e.~which stems from the non-zero twist angle $\theta$ contained in the intra- and inter-layer terms of the twisted TMD Hamiltonian reported in Ref.~\cite{SM})  {\it and} the coupling between spin and orbital motion due the SOC term of the Hamiltonian, allowing us to perform a first exploration of their interplay in a chiral quasi-2D solid-state system.

{\color{blue}{\it Results.}}---We now present our main numerical results.
We used the Kwant software package~\cite{Groth2014} to implement the tight-binding model Hamiltonian introduced above and calculate the spin-resolved $S_{\sigma,\sigma^\prime}$ matrix of the two-terminal 3D transport setup presented in Fig.~\ref{fig1}. Top and bottom leads are modelled as semi-infinite A-A stacked graphite leads. Each graphite lead is rotated in order to be aligned with the adjacent layer belonging to the twisted homobilayer TMD. The leads have Bloch translational invariance both along the $\hat{\bm x}$-$\hat{\bm y}$ plane and the $\hat{\bm z}$ direction. The twisted homobilayer TMD has Bloch translational invariance in the $\hat{\bm x}$-$\hat{\bm y}$ plane. The spin-resolved scattering matrix $S_{\sigma,\sigma^\prime}$ has been computed by carrying out integrals over the 2D moir\'{e} lattice Brillouin zone.

Fig.~\ref{fig2} displays the spin polarization $P_t(E)$, as defined in Eq.~\eqref{eqn:spin_polarization}, of the electrons transmitted from the top to the bottom lead. These results refer to twisted homobilayer ${\rm MoTe}_2$~\cite{Wu_PRL_2019} and a twist angle $\theta = 6.01\degree$. (Additional numerical results for twisted ${\rm MoS}_2$, ${\rm MoSe}_2$, ${\rm WS}_2$, ${\rm WSe}_2$, and ${\rm WTe}_2$ are reported in Sect.~II of Ref.~\cite{SM}.) For the SOC parameters, we have used the values $\lambda_0= 220~{\rm meV}$ and $\lambda_{\rm BR}=0$, which have been taken from Ref.~\cite{Wu_PRL_2019}. We clearly see that the spin polarization exceeds $20\%$ (see also Fig.~\ref{fig3}(a), where much larger values have been obtained for larger twist angles), which is a giant value, given the atomic thickness of the scatterer. As expected, the calculated spin polarization has opposite sign for two opposite twist angles, $\theta$ and $-\theta$. Importantly, the spin polarization vanishes in the non-chiral $\theta=0$ case.

The dependence of the chirality-induced spin polarization $P_t(E)$ on the twist angle $\theta$ is displayed in Fig.~\ref{fig3}(a) for three values of $\theta$. Notice that decreasing $\theta$, from $\theta = 13.17\degree$ down to $\theta = 3.48\degree$, $|P_t(E)|$ decreases. This is not surprising since in the limit $\theta\to 0$ the system becomes non-chiral and $P_t(E)$ vanishes for all values of $E$, as seen in Fig.~\ref{fig2}. For $\theta=13.17\degree$, the spin polarization nearly reaches the impressive value of $60\%$. Fig.~\ref{fig3}(b) shows the dependence of $P_t(E)$ on the magnitude of $\lambda_0$ (while keeping $\lambda_{\rm BR}=0$). We note that increasing SOC leads to an overall increase of $|P_t(E)|$, pointing out the crucial role of this term in the twisted homobilayer TMD Hamiltonian for the emergence of chirality-induced spin polarization.

\begin{figure}[t]
\centering
\begin{tabular}{c}
\begin{overpic}[width=\columnwidth]{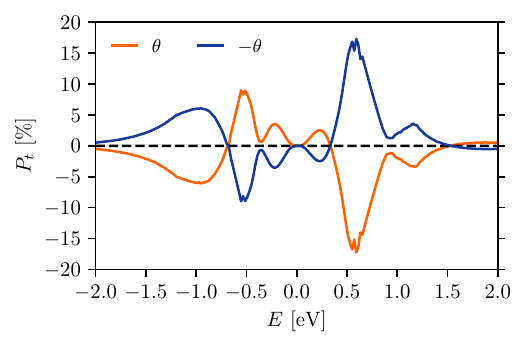} 
\end{overpic}
\end{tabular}
\caption{(Color online)  Same as in Fig.~\ref{fig2}, but for a finite value of $\lambda_{\rm BR}$, $\lambda_{\rm BR} = 50~{\rm meV}$. The black dashed line denotes the results for the non-chiral $\theta=0$ case.\label{fig4}}
\end{figure}

Finally, Fig.~\ref{fig4} shows $P_t(E)$ in the case of finite $\lambda_{\rm BR}$. The important thing to notice is the result for $\theta=0$, i.e.~$P_t(E) =0~\forall E$. At $\theta=0$ and $\lambda_{\rm BR}\neq 0$, our twisted quasi-2D material reduces to a Bernal-stacked homobilayer TMD in the presence of a perpendicular applied electric field. The Hamiltonian of this system breaks inversion symmetry but lacks chirality, yielding {\it no} chirality-induced spin polarization. 

In summary, we have shown that a chirality-induced spin polarization emerges in moir\'{e} materials with strong spin-orbit coupling. Surprisingly, we have found that the spin polarization can be giant even for just two twisted monolayers. We have presented a theory of the effect based on the microscopic Hamiltonian of twisted homobilayer transition metal dichalcogenides~\cite{Wu_PRL_2019}. Numerical results have been obtained for twisted ${\rm MoTe}_2$. In this particular case, the effect appears to be gigantic, provided that the twist angle is sufficiently large. As far as the experimental detection of the effect is concerned, one needs either to employ devices with more than two terminals~\cite{Yang2019} or transcend the linear-response regime~\cite{footnote_van_Wees} (if one insists on using a two-terminal setup). 

During the preparation of this manuscript, we learned about a recent theoretical work reporting chirality-induced spin polarization in bulk (three-dimensional) inorganic crystals with homochiral crystal structures, such as tellurium and transition metal disilicide compounds~\cite{Yang2023}.

{\color{blue}{\it Acknowledgements.---}}This work was partially supported by the European Union's Horizon 2020 research and innovation programme under the Marie Sklodowska-Curie grant agreement No.~873028 - HYDROTRONICS and the FET project QUEFORMAL (Contract No.~829035). M.P., F.T. and G.I. acknowledge support by the MUR through the following PRIN  projects: Project ``q-LIMA'' (Contract. No.~2020JLZ52N), Project ``NEThEQS'' (Contract No.~2022B9P8LN), and Project
``FIVE2D'' (Project No.~$2017{\rm SRYEJH}_001$). F.T. also acknowledges support by the Royal Society through the International Exchanges Scheme between the UK and Italy (Grant No.~IEC R2 192166). P.J.H. acknowledges support by the Gordon and Betty Moore Foundation’s EPiQS Initiative through grant GBMF9463, the Fundacion Ramon Areces, and the ICFO Distinguished Visiting Professor program. F.H.L.K. acknowledges financial support from  the ERC TOPONANOP  (726001), the Government of Catalonia trough the SGR grant, the Spanish Ministry of Economy and Competitiveness through the Severo Ochoa Programme for Centres of Excellence in R\&D (Ref. SEV-2015-0522) and Explora Ciencia (Ref. FIS2017- 91599-EXP), Fundacio Cellex Barcelona, Generalitat de Catalunya through the CERCA program, the Mineco grant Plan Nacional (Ref. FIS2016-81044-P), and the Agency for Management of University and Research Grants (AGAUR) (Ref. 2017-SGR-1656).

We thank P. Novelli and B. van Wees for useful discussions and correspondence.

Computational resources have been provided by {\it computing@unipi}, a computing service provided by University of Pisa.

G.~M., L.~C., and L.~L. and contributed equally to this work.

%Appendix
\clearpage 
\newpage

\setcounter{section}{0}
\setcounter{equation}{0}%
\setcounter{figure}{0}%
\setcounter{table}{0}%

\setcounter{page}{1}

\renewcommand{\thetable}{S\arabic{table}}
\renewcommand{\theequation}{S\arabic{equation}}
\renewcommand{\thefigure}{S\arabic{figure}}
\renewcommand{\bibnumfmt}[1]{[S#1]}
\renewcommand{\citenumfont}[1]{S#1}

\onecolumngrid

\begin{center}
\textbf{\Large Supplemental Material for:\\ ``Chirality-Induced Spin Polarization in Twisted Bilayer Transition Metal Dichalcogenides''}
\bigskip

Guido Menichetti,$^{1}$
Lorenzo Cavicchi,$^{2}$
Leonardo Lucchesi,$^{1,\,3}$
Fabio Taddei,$^{4,\,2}$
Giuseppe Iannaccone,$^{3}$
Pablo Jarillo-Herrero,$^{5}$
Claudia Felser,$^{6}$
Frank H. L. Koppens,$^{7,\,8}$
Marco Polini$^{1,\,7}$

\bigskip

$^1$\!{\it Dipartimento di Fisica dell'Universit\`a di Pisa, Largo Bruno Pontecorvo 3, I-56127 Pisa,~Italy}

$^2$\!{\it Dipartimento di Ingegneria dell'Informazione dell'Universit\`a di Pisa, Via Girolamo Caruso 16, I-56122 Pisa,~Italy}

$^3$\!{\it Scuola Normale Superiore, Piazza dei Cavalieri 7, I-56126 Pisa,~Italy}

$^4$\!{\it NEST, Istituto Nanoscienze-CNR, Piazza S. Silvestro 12, I-56126 Pisa,~Italy}

$^5$\!{\it Department of Physics, Massachusetts Institute of Technology, Cambridge, Massachusetts,~USA}

$^6$\!{\it Max Planck Institute for Chemical Physics of Solids, N\"{o}thnitzer Str. 40, Dresden 01187,~Germany}

$^7$\!{\it ICFO-Institut de Ci\`{e}ncies Fot\`{o}niques, The Barcelona Institute of Science and Technology, Av. Carl Friedrich Gauss 3, 08860 Castelldefels (Barcelona),~Spain}

$^8$\!{\it ICREA-Instituci\'{o} Catalana de Recerca i Estudis Avan\c{c}ats, Passeig de Llu\'{i}s Companys 23, 08010 Barcelona,~Spain}

\bigskip

In this Supplemental Material we present more details on the tight-binding Hamiltonian we have used in our numerical calculations. We also present further numerical results concerning the chirality-induced spin-polarization for two of the most common families of TMDs.
 
\end{center}

\onecolumngrid

\appendix
\section*{Section I: Model hamiltonian}\label{supplementary}
Here we present more details on the tight-binding model we have used to describe electrons roaming in the twisted TMD homobilayer moir\'e superlattice.

We start from the tight binding model for TBG~\cite{supp_Santos2007,supp_Laissardiere2012,supp_Novelli2020}. The basis of Bloch states is built from the $p_z$ atomic orbitals of Carbon. We introduce the localized atomic orbitals centered at the point $\bm{d}_{\tau,\ell} + \bm{t}_{n,\ell}$, i.e.
\begin{equation}
\langle\bm{r}|n,\ell,\tau\rangle = \phi(\bm{r} - \bm{d}_{\tau,\ell} - \bm{t}_{n,\ell})~,
\end{equation}
where $\phi(\bm{r})$ is the wavefunction of a $p_z$ orbital centered at the origin, $\bm{d}_{\tau,\ell}$ is the basis vector of the sublattice $\tau$ in layer $\ell$, whereas the symbol $\bm{t}_{n,\ell}$ is a shorthand for
\begin{equation}
 \bm{t}_{n,\ell} = n_{1}\tilde{\bm{t}}_{1,\ell} + n_{2}\tilde{\bm{t}}_{2,\ell} \quad \text{with } n_{1}, \, n_{2} \in \mathbb{N}~.
\end{equation}
The vectors $\tilde{\bm{t}}_{1/2,\ell}$ are primitive translation vectors of the graphene lattice in layer $\ell$,
and the sum over $n$ should be intended as
\begin{equation}
  \sum_{n} [\cdots] = \sum_{n1,n2 \in \mathbb{N}} [\cdots]~.
\end{equation}
The atomic orbitals are assumed to be orthogonalized according to 
\begin{equation}
 \langle n,\ell,\tau |n^\prime,\ell^\prime,\tau^\prime\rangle= \delta_{n,n^{\prime}}\delta_{\ell,\ell^{\prime}}\delta_{\tau,\tau^{\prime}}~.
\end{equation}
In the two-center approximation, and retaining only the nearest-neighbour contributions, the intra-layer Hamiltonian of graphene in layer $\ell$ takes the form
\begin{equation}\label{eq:app_intra_layer_hamiltonian}
  \hat{H}^{(\ell)}_{{\rm intra}} = -t \sum_{\langle m,n \rangle}\sum_{\tau,\tau^{\prime}} |m,\ell,\tau\rangle\langle n, \ell, \tau^{\prime}|(1 - \delta_{\tau,\tau^{\prime}})~,
\end{equation}
where the energy $t$ is given by
\begin{equation}
  -t \equiv \int d\bm{r}~ \phi^{*}(\bm{r})V(\bm{r}-\bm{d}_{\tau,1})\phi(\bm{r} - \bm{d}_{\tau,1}) = \int d\bm{r}~ \phi^{*}(\bm{r})V(\bm{r}-\bm{d}_{\tau,2})\phi(\bm{r} - \bm{d}_{\tau,2})~,
\end{equation}
$V(\bm{r})$ being the spherically-symmetric potential of a Carbon atom centered at the origin. The sum over $\langle m,n\rangle$ runs over neighboring orbitals, i.e. the states $|m,\ell,\tau\rangle$ and $|n,\ell,\tau^{\prime}\rangle$ in Eq.~\eqref{eq:app_intra_layer_hamiltonian} correspond to neighbouring orbitals. 

In this work we have chosen the following primitive translation vectors
\begin{equation}
\tilde{\bm{t}}_{1/2,\ell=1} = \left(\mp\frac{a}{2}, \frac{a\sqrt{3}}{2}\right)~, \quad \tilde{\bm{t}}_{1/2,\ell=2} = {\cal R}(\theta)\left(\mp\frac{a}{2}, \frac{a\sqrt{3}}{2}\right)~,
\end{equation}
where ${\cal R}(\theta)$ is the rotation matrix defined by:
\begin{equation}
    {\cal R}\left(\theta\right) = \cos(\theta) \mathbb{I}_{2\times2} - i\sin(\theta)\sigma_y = \left(\begin{array}{cc}
        \cos(\theta) & \pm\sin(\theta) \\
        \sin(\theta) & \cos(\theta)
    \end{array}\right)~.
\end{equation}
As in the main text, layer 1 (top) is the reference layer, while layer 2 (bottom) is counter-clockwise rotated by an angle $\theta$. In addition, the basis vectors are
\begin{equation}\label{eq:AB-stacking}
    \bm{d}_{\tau,\ell} =
    \begin{cases}
        \frac{a}{\sqrt{3}} \left( -\frac{\sqrt{3}}{2},\frac{1}{2}\right)~,& \text{if layer} = 1 \text{ and sub-lattice} = B~.\\
        - \frac{a}{\sqrt{3}}  {\cal R}(\theta)\left( -\frac{\sqrt{3}}{2},\frac{1}{2}\right)~,& \text{if layer} = 2 \text{ and sub-lattice} = A~.\\
        \bm{0}~,              & \text{otherwise~.}
    \end{cases}
\end{equation}
The choice of these translation and basis vectors is such that in the limit $\theta\rightarrow 0 $ one obtains AB-stacked bilayer graphene.

We now describe the tunneling of electrons between orbitals in different layers. The inter-layer Hamiltonian can be written as
\begin{equation}
  \hat{H}_{{\rm inter}}^{(\ell,\ell^{\prime})} = \sum_{n, n^{\prime}}\sum_{\tau,\tau^{\prime}} h_{\tau,\tau^{\prime}}(\bm{d}_{\tau,\ell} + \bm{t}_{n,\ell} - \bm{d}_{\tau^{\prime},\ell^{\prime}} - \bm{t}_{n^{\prime},\ell^{\prime}})|n,\ell,\tau\rangle \langle n^{\prime},\ell^{\prime},\tau^{\prime}| + {\rm H.c.}~.
\end{equation}
An empirical form of the transfer integral between two $p_z$ orbitals of Carbon atoms in the Slater-Koster approximation is given by \cite{supp_Laissardiere2012}:
\begin{equation}\label{eqn:inter-layer}
   h_{\tau,\tau^{\prime}}(\bm{R}_{n,n^\prime}) = -t \exp\left(-\frac{|\bm{R}_{n,n^\prime}| - a}{\lambda}\right) \frac{|\bm{R}_{n,n^\prime}\cdot\bm{e}_{\parallel}|^2}{|{\bm R}_{n,n^\prime}|^2} + t_{\perp}\exp\left(-\frac{|\bm{R}_{n,n^\prime}| - d}{\lambda}\right) \frac{|\bm{R}_{n,n^\prime}\cdot\bm{e}_{\perp}|^2}{|{\bm R}_{n,n^\prime}|^2}~.
\end{equation}
In Eq.~(\ref{eqn:inter-layer}) we have introduced the short-hand notation $\bm{R}_{n,n^\prime}\equiv \bm{d}_{\tau,\ell} + \bm{t}_{n,\ell} - \bm{d}_{\tau^{\prime},\ell^{\prime}} - \bm{t}_{n^{\prime},\ell^{\prime}}$, for which indeed $h_{\tau,\tau^{\prime}}(\bm{d}_{\tau,\ell} + \bm{t}_{n,\ell} - \bm{d}_{\tau^{\prime},\ell^{\prime}} - \bm{t}_{n^{\prime},\ell^{\prime}}) = h_{\tau,\tau^{\prime}}(\bm{R}_{n,n^\prime})$. In Eq.~\eqref{eqn:inter-layer} we have also introduced the decay length $\lambda$ and the inter-layer distance $d$. The hopping parameter $t_{\perp}$ is given by:
\begin{equation}
   t_\perp = \int d\bm{r} \phi^*(\bm{r}) V(\bm{r} - d\bm{e}_z)\phi(\bm{r} - d\bm{e}_z)~,
\end{equation}
which is the transfer integral between two $p_z$ orbitals that are one on top of the other, vertically displayed by a distance $d$.

In this work we study vertical transport through twisted homobilayer TMDs, whose Hamiltonian can be obtained by modifying the above tight-binding model into a twisted bilayer gapped graphene one~\cite{supp_Wu_PRL_2019}. To this end, we reproduce the monolayer TMD energy gap $\Delta$ and carrier effective mass $m^\ast$ by adding a staggered potential $E_{A},E_{B}$ on the sub-lattice sites and re-scaling the intra-layer nearest-neighbor hopping $t$, while retaining the lattice parameter of graphene $a$:
\begin{equation}
   \hat{H}^{(\ell)}_{\rm intra} = \sum_{m}\sum_{\tau} E_{\tau}|m, \ell, \tau\rangle\langle m, \ell, \tau| - \tilde{t} \sum_{\langle m, n \rangle}\sum_{\tau\tau^\prime} |m, \ell, \tau\rangle\langle n, \ell, \tau^\prime| (1 - \delta_{\tau\tau^\prime})~,
\end{equation}
where
\begin{equation}
\tilde{t}=\frac{\hbar}{\sqrt{3}a}\sqrt{\frac{2\Delta}{3m^\ast}}~.
\end{equation}
The numerical results presented and discussed in the main text are obtained for the following choice of parameters: $a=0.246$~nm, $\Delta = 1.1~{\rm eV}$, and $m^\ast = 0.6~m_{\rm e}$, $m_{\rm e}$ being the bare electron mass in vacuum. The inter-layer hopping energy and distance appearing in the inter-layer Hamiltonian term~\eqref{eqn:inter-layer} are instead fixed to $t_\perp = 0.15$~eV and $d = 0.69$~nm, while the exponential decay length is kept equal to the one in TBG, i.e.~$\lambda = 0.184 \sqrt{3} a$~\cite{supp_Laissardiere2012}. These parameters have been chosen in order to effectively reproduce the energy bands of twisted MoTe$_2$. 

\begin{figure}[t]
\centering
\begin{tabular}{cc}
\begin{overpic}[width=0.5\columnwidth]{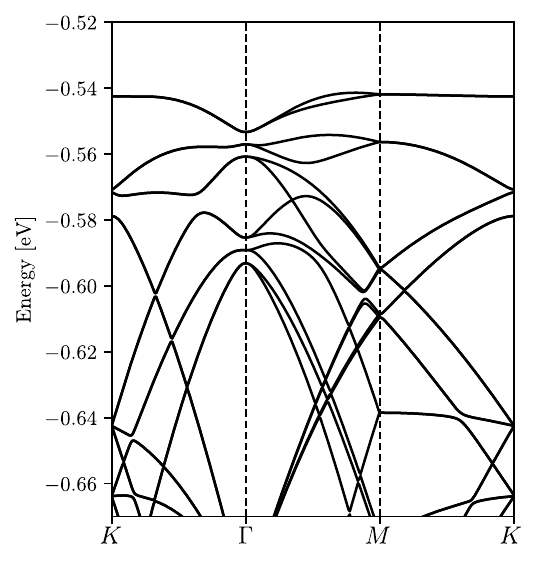}
\put(1,95){\rm{(a)}}
\end{overpic} &
\begin{overpic}[width=0.5\columnwidth]{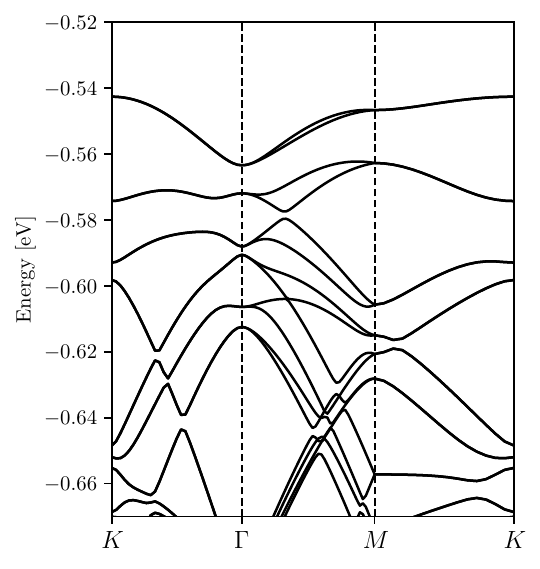}
\put(1,95){\rm{(b)}}
\end{overpic}
\end{tabular} 
\caption{Band structure of twisted ${\rm MoTe}_2$ obtained from our tight-binding model---panel (a)---and from the  continuum model of Ref.~\cite{supp_Wu_PRL_2019}---panel (b). Results in this plot refer to a twist angle $\theta = 3.89\degree$. The parameters of the continuum model Hamiltonian are extracted from Ref.~\cite{supp_Wu_PRL_2019}. The dependence of the bands on the wave-vector ${\bm k}$ is displayed along the high-symmetry path $K$-$\Gamma$-$M$-$K$ of the moir\'{e} Brillouin zone.\label{figS1}}
\end{figure}

In Fig.~\ref{figS1} we compare our results for the energy bands obtained from the above described tight-binding model---panel (a)---with the results of Ref.~\cite{supp_Wu_PRL_2019}, which were obtained from a continuum model Hamiltonian.
\section*{Section II: Additional numerical results}\label{supplementary_NR}
\begin{figure}[h]
\centering
\begin{tabular}{cc}
\begin{overpic}[width=0.5\columnwidth]{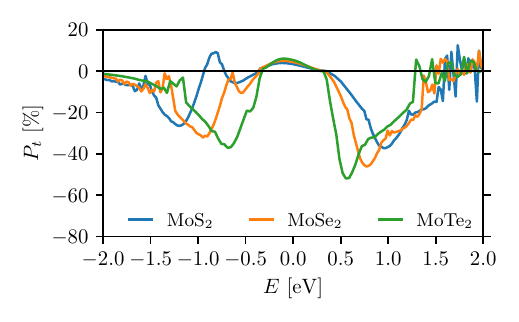}
\put(1,56){\rm{(a)}}
\end{overpic} &
\begin{overpic}[width=0.5\columnwidth]{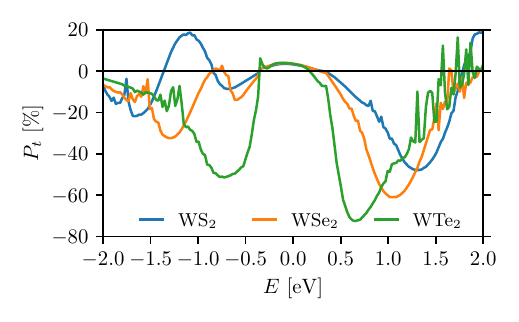}
\put(1,56){\rm{(b)}}
\end{overpic}
\end{tabular} 
\caption{(Color online) Results for the spin polarization $P_t(E)$ for two different families of twisted TMDs. Results in this figure refer to a single twist angle,~i.e.~$\theta = 13.17\degree$. Panel (a) Results for ${\rm MoX}_2$ with ${\rm X} = {\rm S}, {\rm Se}$, and ${\rm Te}$. Panel (b) Results for ${\rm WX}_2$ with ${\rm X} = {\rm S}, {\rm Se}$, and ${\rm Te}$. In both panels the solid black line has been added as a guide to the eye. The parameters of the microscopic Hamiltonians used to produce these numerical results have been reported in Table~\ref{tableS1}. \label{figS2}}
\end{figure}

In this Section we present additional numerical results. In the main text, indeed, we focussed on for the case of twisted ${\rm MoTe}_2$. For the sake of completeness, in Fig.~\ref{figS2} we show the chirality-induced spin polarization $P_t$ for two of the most common families of TMDs. In panel (a) we report results for twisted homobilayers composed by ${\rm MoS}_2$, ${\rm MoSe}_2$, and ${\rm MoTe}_2$ (already presented in the main text). In panel (b) we do the same for ${\rm WS}_2$, ${\rm WSe}_2$, and ${\rm WTe}_2$. Both panels have been calculated by setting $\theta=13.17^\circ$.

The microscopic parameters of the Hamiltonian used to compute the spin polarization $P_t(E)$ in Fig.~\ref{figS2} are collected in Table~\ref{tableS1}. Each Hamiltonian contain the following parameters: the energy gap $\Delta$, the hole-carrier effective mass $m^\ast$ (expressed in units of the bare electron mass in vacuum), the SOC parameter $\lambda_0$ discussed in the main text, and the interlayer distance $d$. The Table reports also the SOC to energy gap ratio $\lambda_0/\Delta$ and the maximum of the absolute value of the chirality-induced spin-polarization:
\begin{equation}\label{eq_ptmax}
|P_t|^{\rm max} \equiv  \max_{E\in [-2,2]{\rm eV}} |P_t(E)|~, 
\end{equation}
extracted from Fig.~\ref{figS2}. As it can be inferred from Fig.~\ref{figS2}, within a given TMD family, $|P_t|^{\rm max}$ is achieved in the material with the maximum value of $\lambda_0/\Delta$.

\begin{table}[h]
    \centering
    \begin{tabular}{|l|l|l|l|l|l|l|}
         \hline 
         & $\Delta~[{\rm eV}]$ & $m^\ast~[{m_{\rm e}}]$  & $\lambda_0~[{\rm meV}]$ & $d$~[{\rm nm}]~& $\lambda_0/\Delta$ ~& $|P_t|^{\rm max}~[{\rm \%}]$   \\
         \hline
        MoS$_2$~\cite{supp_Zibouche_2014,supp_Kim_2021} & 1.62 & 0.58	& 255	& 0.62 & 0.16	&37.3 \\ \hline
        MoSe$_2$~\cite{supp_Zibouche_2014,supp_Kim_2021}& 1.4	&0.67	&	263&	0.65& 0.19	&46.2 \\ \hline
        MoTe$_2$~\cite{supp_Wu_PRL_2019} & 1.1 & 0.62	&	220	& 0.69& 0.20	& 59.5 \\ \hline
        WS$_2$~\cite{supp_Zibouche_2014,supp_Kim_2021} &1.74	&0.42	&	459&	0.61	& 0.26 &48.0 \\ \hline
        WSe$_2$~\cite{supp_Zibouche_2014,supp_Kim_2021} &1.43	&0.45	&	446&	0.64 & 0.31	&61.1 \\ \hline
        WTe$_2$~\cite{supp_Zibouche_2014,supp_Kim_2021} &0.86	&0.41	&	401&	0.69	& 0.47 & 72.5 \\ \hline
    \end{tabular}
    \caption{Numerical parameters employed to obtain the results shown in Fig.~\ref{figS2}. Energy gap $\Delta~({\rm eV})$, hole-carrier effective mass  $m^\ast~({m_{\rm e}})$ in units of the bare electron mass in vacuum $m_{\rm e}$, the SOC parameter $\lambda_0$ discussed in the main text, the interlayer distance $d$ (in ${\rm nm}$), the SOC to energy gap ratio $\lambda_0/\Delta$, and $|P_t|^{\rm max}$ as defined in Eq.~(\ref{eq_ptmax}). This last quantity has been extracted from Fig.~\ref{figS2} and refers to $\theta=13.17^\circ$.}
    \label{tableS1}
\end{table}

\begin{figure}[h]
\centering
\begin{tabular}{c}
\begin{overpic}[width=0.5\columnwidth]{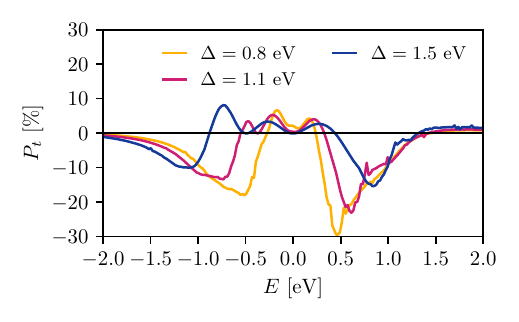}
\end{overpic}
\end{tabular} 
\caption{(Color online) Chirality-induced spin polarization $P_t(E)$ in ${\rm MoTe}_2$.  Different colors refer to different values of $\Delta$: $\Delta = 0.8~{\rm eV}$ (orange), $\Delta = 1.1~{\rm eV}$ (magenta), and $\Delta = 1.5~{\rm eV}$ (blue). All the results have been obtained by setting $\theta=6.01\degree$.\label{figS3}}
\end{figure}

In order to illustrate the role of the energy gap $\Delta$ on the quantity $P_t(E)$, in Fig.~\ref{figS3} we plot three different results for $P_t(E)$  in ${\rm MoTe}_2$. Different curves in this figure have been obtained by artificially changing the energy gap $\Delta$ while keeping fixed all the other microscopic parameters of ${\rm MoTe}_2$. The twist angle is fixed at $\theta=6.01\degree$. It is evident that $|P_t|^{\rm max}$ increases with decreasing $\Delta$ (thereby increasing the value of the ratio $\lambda_0/\Delta$).

\begin{figure}[t]
\centering
\begin{tabular}{c}
\begin{overpic}[width=0.5\columnwidth]{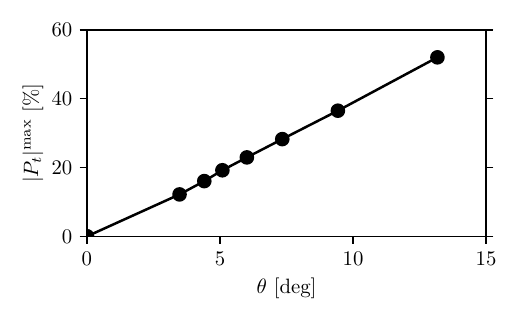}
\end{overpic}
\end{tabular} 
\caption{The quantity $|P_t|^{\rm max}$ is plotted as a function of the twist angle $\theta$. The black solid line is a guide to the eye. All the other parameters are the same as in Fig.~\ref{fig2} in the main text.\label{figS4}}
\end{figure}

We conclude this Section by showing the dependence of $|P_t|^{\rm max}$ on the twist angle $\theta$ in Fig.~\ref{figS4}. Results in this plot refer to twisted ${\rm MoTe}_2$. For the computed angles, a linear behavior is evident.

\end{document}